THE CURIOUS CASE OF LEMAÎTRE'S EQUATION NO. 24

Sidney van den Bergh
Dominion Astrophysical Observatory, Herzberg Institute of Astrophysics, National Research Council of Canada, 5071 West Saanich Road, Victoria, British Columbia, V9E 2E7, Canada,
sidney.vandenbergh@nrc.ca

In August of 1961 Abbé Georges Lemaître told me (with a twinkle in his eyes) that, being a priest, he felt a slight bias in favor of the idea that the Universe had been created. It must therefore have been a particular pleasure for him (Lemaître 1927) to have been the first to find both observational and theoretical evidence for the expansion of the Universe. His observational discovery was based on the published distances and radial velocities of 42 galaxies. Lemaître's theretical result was based on the finding that the Universe is unstable, so that pertubations tend to grow. These results, which were published in French and in a relatively obscure journal, anticipated the work of Edwin Hubble (1929) by two years. It might therefore have been appropriate to assign the credit for the discovery of the expansion of the Universe to Lemaître, rather than to Hubble (Peebles 1984). The early evolution of our understanding of the expansion (and the scale-size) of the Universe has recently been discussed in detail by Kragh & Smith (2003) and by Nussbaumer & Bieri (2009).

Because it had been published in such a low-impact place an authorized translation of Lemaître's discovery paper was reprinted in the widely read Monthly Notices of the Royal Astronomical Society (Lemaître 1931). It is this translation, rather than the French original, which formed the basis of most subsequent discussions of the discovery of the expansion of the Universe. A comparison between the original French text and its English translation shows a few, but very interesting, differences (e.g. Peebles 1984, Way & Nussbaumer 2011). It does not previously seem to have been noted that one of the 31 equations in Lemaître's paper is also different in the original and in its translation. [The unknown translator did his work well and corrected a typographical error in one of these equations in the French original version of the paper.] In the English translation the term

$$\frac{v}{rc} = \frac{625 \; x \; 10^5}{10^6 x \; 3{,}08 \; x \; 10^{18} \; x \; 3 \; x \; 10^{10}},$$

in which v is the radial velocity, r the distance and c the velocity of light is omitted. Of the three numbers given above the speed of light in cm/s, and the length of the parsec are well-known. Only the cosmic expansion term $625 \; x \; 10^5$ [corresponding in modern parlance to a Hubble constant of 625 km/s] might possibly be considered to be controversial. The fact that dropping this term from Lemaître's Eqn. 24 was intentional is supported by the fact that a short paragraph in the paper, which deals with the determination of what we now call the Hubble parameter was also omitted from the English translation of the text. (The latter fact had already been noted previously by Peebles (1984) and by Way & Nussbaumer 2011). That mention of the expansion of the Universe was omitted from the English version of both Eqn. 24, and from the English text, suggests that this exclusion by the translator was deliberate rather than accidental. The Editor-in-chief of the Monthly Notices has kindly informed me that his office no longer has any records of the events related to the translation of Lemaître's article in 1931. Another factor which may have influenced the lack of credit assigned to Lemaître for the discovery of the expansion of the Universe is that the English translation of the article did not include the footnotes to the original French version of the article. One of these footnotes explains in detail how using weighted and unweighted radial velocities for galaxies leads to slightly different values for the Hubble parameter. In summary it appears that the translator of Lemaître's 1927 article deliberately deleted those parts of the paper that dealt with the determination of what is presently referred to as the Hubble parameter. The reason for this remains a mystery.

I am indebted to Bob Carswell for information on the files of the Monthly Notices and to Harry Nussbaumer for helpful exchanges on early work related to the expansion of the Universe.